\documentclass[12pt]{iopart}
\usepackage{graphicx}
\font\tenimbf=cmmib10 at 10pt
\font\sevenimbf=cmmib10 at 6pt
\font\fiveimbf=cmmib10 at 4pt
\newfam\imbf
\textfont\imbf=\tenimbf
\scriptfont\imbf=\sevenimbf
\scriptscriptfont\imbf=\fiveimbf

\def\bea{\begin{eqnarray}}
\def\eea{\end{eqnarray}}  
\def\beq{\begin{equation}}
\def\eeq{\end{equation}}  
\begin{document}

\title[Antibaryon to Baryon Production Ratios]{Antibaryon to Baryon
Production Ratios\\
 in Pb-Pb and p-p collision at LHC energies\\
 of the DPMJET-III Monte Carlo}

%
%

\author{F.Bopp$^1$, R.Engel$^2$,  J.Ranft$^1$,    and
S.Roesler$^3$}

\address{$^1$ Siegen University, Siegen, Germany}
\address{$^2$ Forschungszentrum Karlsruhe, Karlsruhe, Germany}
\address{$^3$ CERN, Geneva, Switzerland}
\ead{bopp@physik.uni-siegen.de}
\begin{abstract}
A sizable component of stopped baryons is predicted for $pp$ and
$PbPb$ collisions at LHC. Based on an analysis of RHIC data within
framework of our multichain Monte Carlo DPMJET-III the LHC predictions
are presented. 
\end{abstract}

\maketitle

\begin{flushleft}{ This addendum to Ranft's talk about the main
DPMJET III prediction addresses baryon stopping. The interest is a
component without leading quarks. Where the flavor decomposition is
not determined by final state interactions the valence-quarkless component
can be enhanced by considering net strange baryons. }\end{flushleft}{ \par}

{ \par{}}{ \par}

\begin{flushleft}{ In models, in which soft gluons can arbitrarily
arrange colors, a configuration can appear in which the baryonic charge
ends up moved to the center. The actual transport is just an effect
of the orientation of the color-compensation during the soft hadronisation.
Various other ideas about fast baryon stopping exist but to have it
caused by such an {}``initial'' process is an attractive option.
\par{}}\end{flushleft}{ \par}

\begin{flushleft}{ The {}``Dual-Topological'' phenomenology
of such baryon transport processes was developed 30 years ago\cite{venez77}.
Critical are various baryonium-exchange intercepts which were  estimated
at that time. Some ambiguity remains until today for the quarkless
com\-po\-nent (also called ``string junction" exchange denoted as $\{SJ\}$)  
and a confirmation of the flat net-baryon distribution indicated
by RHIC data at LHC would be helpful.\par{}}\end{flushleft}{ \par}

\begin{flushleft}{ Nowadays it is postulated that at very high
energy hadronic scattering can be understood as extrapolation of BFKL
Pomeron exchanges\cite{lipatov99} and their condensates in the minimum
bias region. BFKL Pomerons are described by ladders of dispersion
graphs, in which soft effects are included using effective gluons.
In principle these soft effects include the color compensating mechanism
usually modelled as two strings neutralizing triplet colors.} A necessary
ingredient in this approach are \emph{Odderons} exchanges with Pomeron-like
intercepts and with presumably much smaller couplings. As these Odderons
can produce a baryon exchange of the type discussed above, a small
rather flat net baryon component is expected.{ \par{}}\end{flushleft}{ \par}

\begin{flushleft}Experimentally, the first indication for a flat component
came from never finalized preliminary ZEUS data at HERA. As RHIC runs
$pp$ or \emph{heavy ions} instead of $p\bar{p}$ this question could
be addressed much better than before and the data seem to require
a flat contribution. In a factorizing Quark-Gluon-String model calculation\cite{Bopp:2006dv}
the best fit { to RHIC BRAHMS $pp$ data at $\sqrt{{s}}=200$
GeV required diquarks with a probability of $\epsilon=0.024$ to involve
a quarkless baryonium-exchange with an intercept $\alpha_{\{ SJ\}}=0.9$.}
{ \par{}}\end{flushleft}{ \par}

\begin{flushleft}To obtain such a { quarkless baryonium-exchange
in the microscopic generator DPMJET III\cite{dpmjet2} a} new
string interaction { reshuffling the initial strings was introduced.
It introduces an exchange with a conservative intercept of $\alpha_{\{ SJ\}}=0.5$.
With this baryonium addition good fits were obtained for various baryon
ratios in} $p-p$ { and $d-Au$ RHIC and $\pi-p$ FERMILAB
processes\cite{Bopp:2005cr}.
There are of course a number of more conventional baryon transport
mechanisms implemented in the model. As the string interaction requires
multiple Pomeron exchanges the new mechanism is actually only a $10$\%
effect at} $pp$ { RHIC. It is, however, important for heavy
ion scattering or at LHC energies. }\end{flushleft}{ \par}

{ \par}{ \par}

\begin{flushleft}\begin{minipage}[b]{6cm}\begin{flushleft}For \emph{}\(pp\) LHC the DPMJET III prediction for
the pseudo rapidity of \(p\), \(\bar{p}\), and \(p-\bar{p}\)
is shown in the Figure on the right. The new baryon stopping is now 
a 40\% effect\emph{.} Of course, with the effective
intercept of \(0.5\) the present implementation of the baryon stopping
is a rather conservative estimate. For an intercept of \(1.0\) the
value at \(\eta=0\) would roughly correspond to the present value of
\(\eta=4\)\end{flushleft}\end{minipage}\vspace*{2cm} 
\includegraphics[scale=0.50]{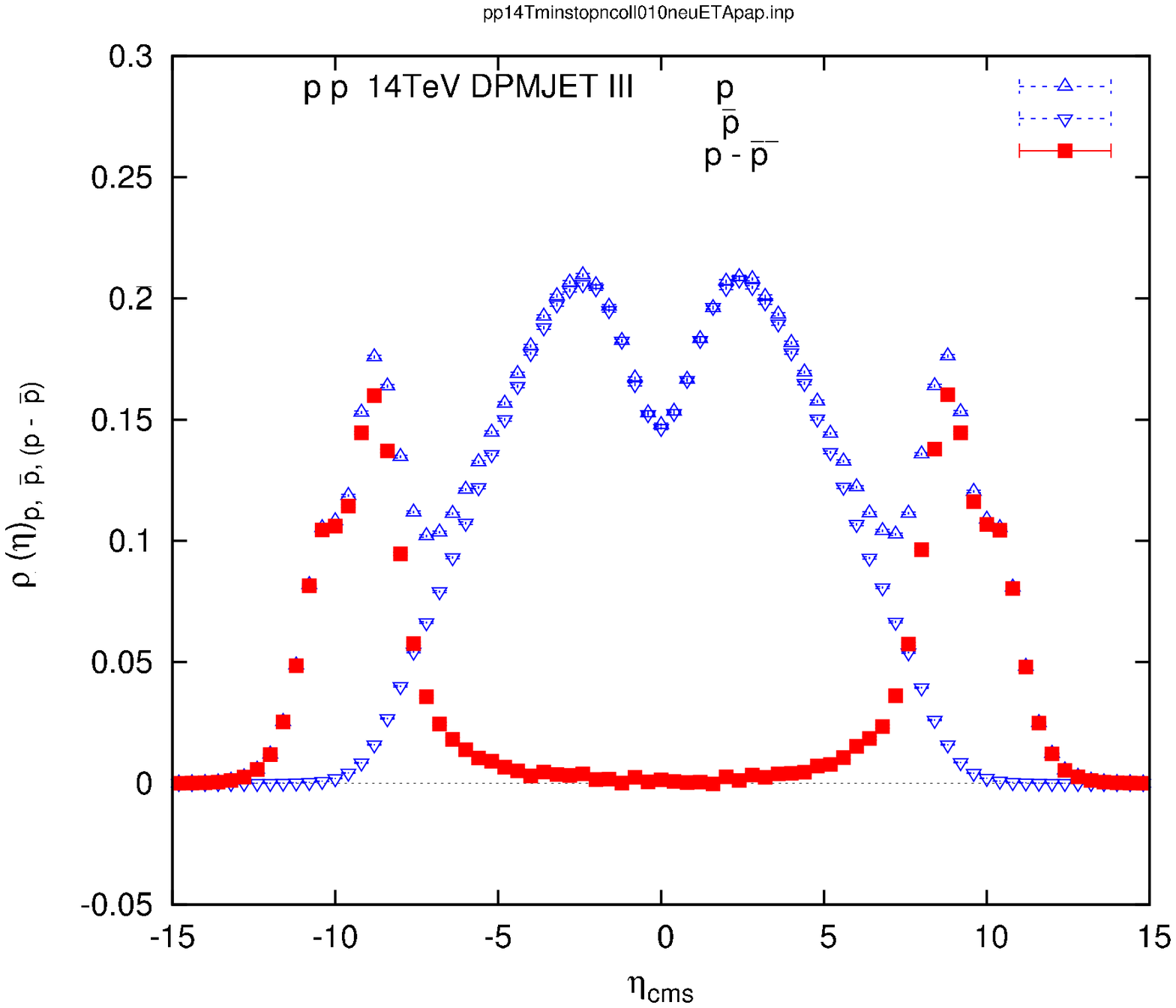}\end{flushleft}\vspace*{-1.5cm}

\noindent
\begin{minipage}[b]{11cm}
\begin{flushleft} \includegraphics[scale=0.50]{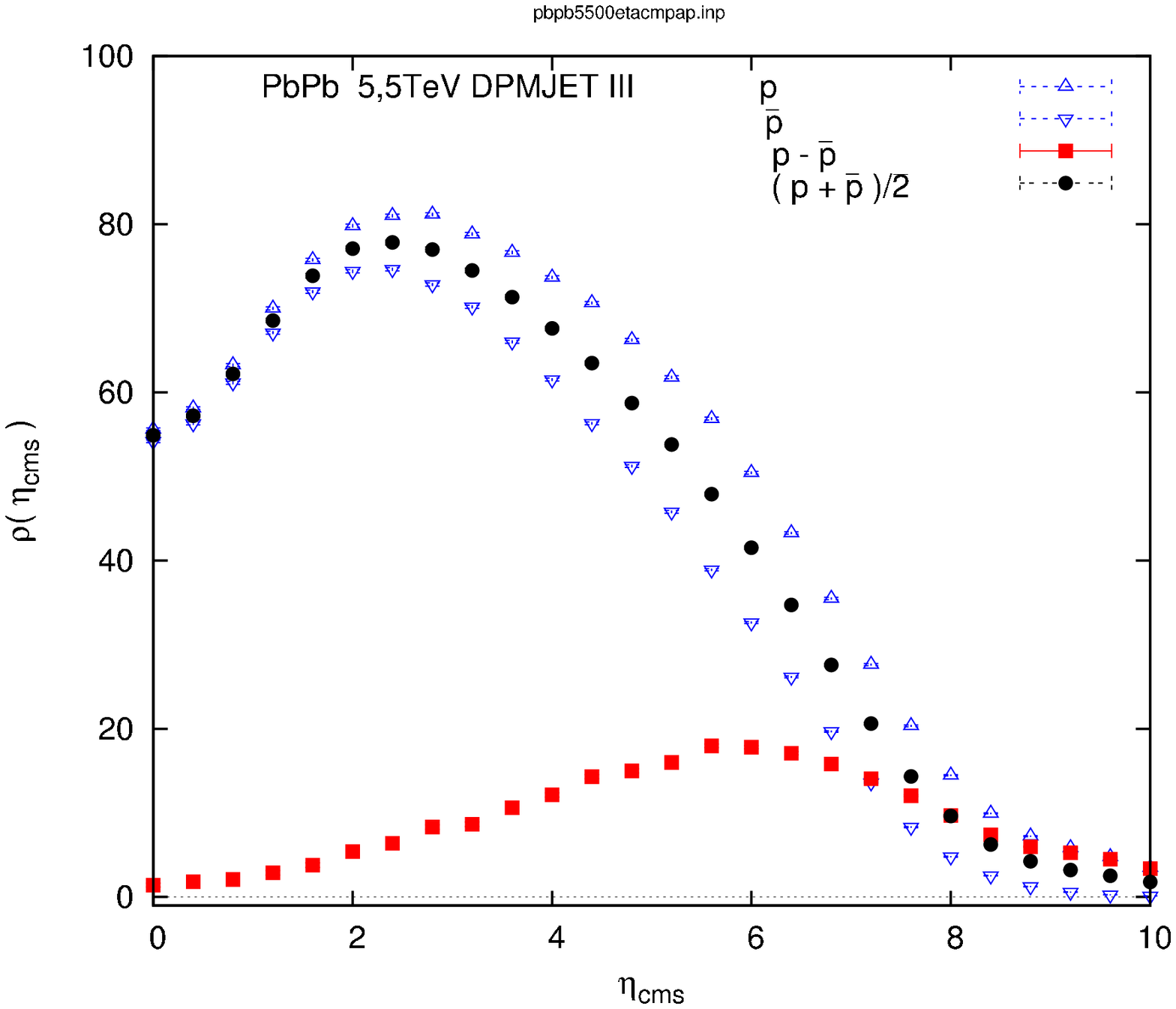}
\end{flushleft} 

\end{minipage}    \begin{minipage}[b]{4cm}
We now turn to DPMJET III prediction for central $PbPb$ LHC. For
the most central $10\%$ of the heavy ion events the pseudorapidity
proton and $\Lambda$ distributions are given in figures below. The
$PbPb$ results are preliminary, as the model is not well tested
in this region.\vspace{-1.5cm}\end{minipage}
\vspace{-0.2cm}

\noindent
\begin{minipage}[b]{11cm}
   \begin{flushleft}
\includegraphics[scale=0.50]{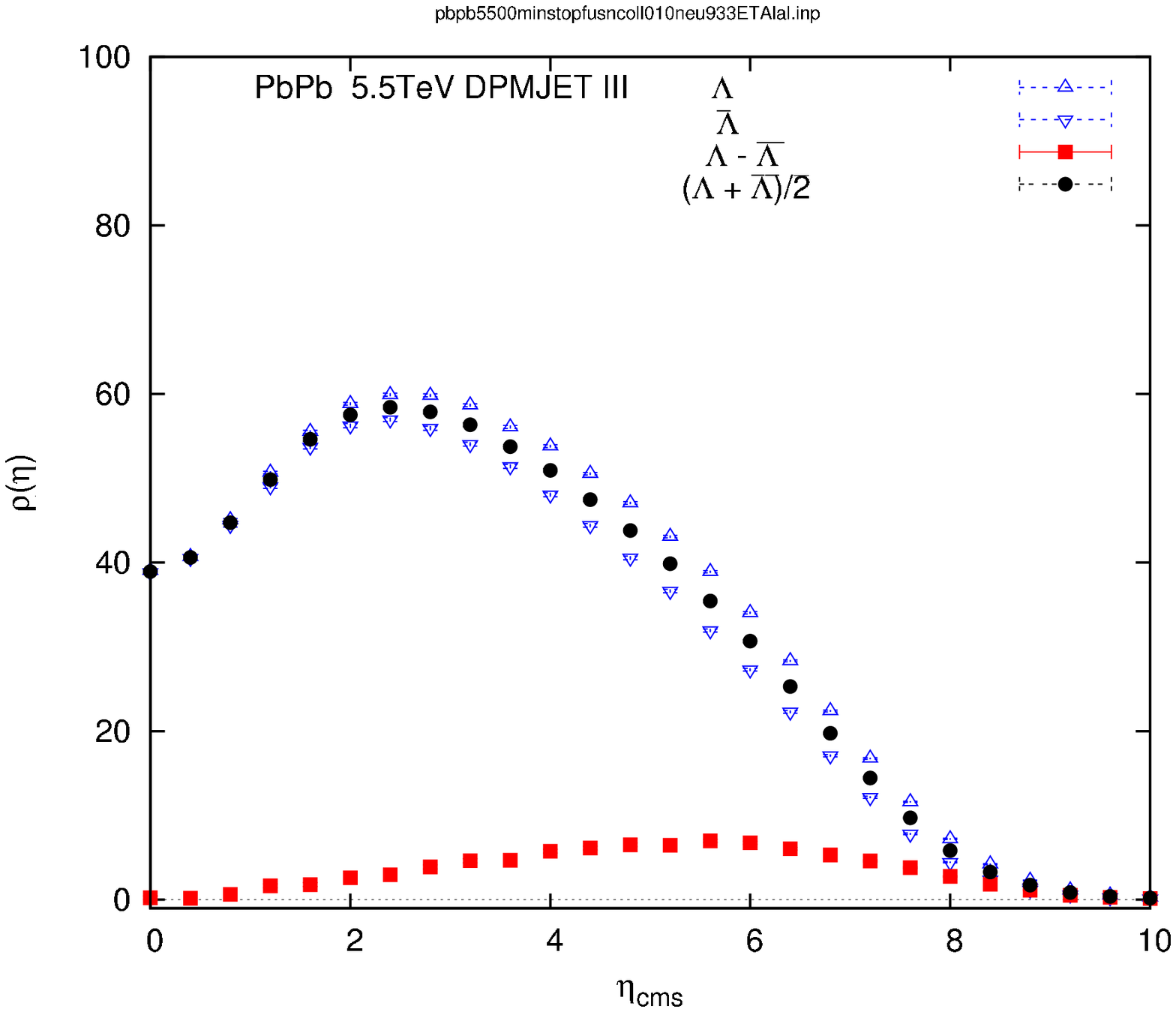}\end{flushleft}

\end{minipage}   \begin{minipage}[b]{4cm}$  $ \end{minipage}

\end{document}